\documentclass[12pt]{article}
\usepackage{epsfig}
\usepackage{amsfonts}
\usepackage{latexsym}
\usepackage{amsmath,amssymb}
\usepackage{verbatim}
\usepackage{setspace}
\usepackage[textheight=9in, textwidth=6.5in, letterpaper]{geometry}
\def\half{{1\over 2}}
\numberwithin{equation}{section}

\def\calr{{\cal R}}

 \def\p{\partial}

\def\co{{\cal O}}
\newcommand{\bea}{\begin{eqnarray}}
\newcommand{\eea}{\end{eqnarray}}
\newcommand{\be}{\begin{equation}}
\newcommand{\ee}{\end{equation}}
\newcommand{\ba}{\begin{align}}
\newcommand{\ea}{\end{align}}

\renewcommand{\O}{\Omega}

\newcommand{\tr}{\mbox{tr}}

\renewcommand{\epsilon}{\varepsilon}

  \makeatletter
  \let\over=\@@over \let\overwithdelims=\@@overwithdelims
  \let\atop=\@@atop \let\atopwithdelims=\@@atopwithdelims
  \let\above=\@@above \let\abovewithdelims=\@@abovewithdelims

\renewcommand\section{\@startsection {section}{1}{\z@}%
                                   {-3.5ex \@plus -1ex \@minus -.2ex}%nn
                                   {2.3ex \@plus.2ex}%
                                   {\normalfont\large\bfseries}}

\renewcommand\subsection{\@startsection{subsection}{2}{\z@}%
                                     {-3.25ex\@plus -1ex \@minus -.2ex}%
                                     {1.5ex \@plus .2ex}%
                                     {\normalfont\bfseries}}

\def\half{{1 \over 2}}

\def\tr{{\rm Tr}}

\def\Or[#1]{{\text{O}}\left({#1}\right)}
\def\dotl[#1,#2]{\left\langle #1, #2 \right\rangle}
\def\dotlb[#1,#2]{[ #1, #2 ]}
\def\dotp[#1,#2]{(#1) \cdot (#2)}
\def\aff[#1,#2]{\hat{#1}(#2)}
\def\n4sym{{\cal N}=4 SYM}
\def\>{\rangle}
\def\<{\langle}
\def\weight[#1,#2,#3]{\{(#1),#2,#3\}}
\def\ads[#1]{$\text{AdS}_{#1}$}

\begin{document}
\begin{titlepage}
\unitlength = 1mm
\ \\

\begin{center}

{ \LARGE {\textsc{State/Operator Correspondence in Higher-Spin DS/CFT}}}

\vspace{0.8cm}
Gim Seng Ng and Andrew Strominger

\vspace{1cm}

{\it  Center for the Fundamental Laws of Nature, Harvard University,\\
Cambridge, MA 02138, USA}

\begin{abstract}
 A recently conjectured microscopic realization of the dS$_4$/CFT$_3$ correspondence
relating Vasiliev's higher-spin gravity on dS$_4$ to a Euclidean $Sp(N)$ CFT$_3$  is used to illuminate some previously inaccessible aspects of the dS/CFT dictionary. In particular it is
argued that states of the boundary CFT$_3$ on $S^2$ are holographically dual to bulk states on geodesically complete,  spacelike $R^3$ slices which terminate on an $S^2$ at future infinity.
The dictionary is described in detail for the case of free scalar excitations.  The ground states of the free or critical $Sp(N)$ model are dual to
dS-invariant plane-wave type vacua, while the bulk Euclidean vacuum is dual to a certain mixed state in the CFT$_3$.  CFT$_3$ states created by operator insertions are found to be dual to (anti) quasinormal modes in the bulk.  A norm is defined on the $R^3$ bulk Hilbert space and shown for the scalar case to be equivalent to both the Zamolodchikov and pseudounitary C-norm of the  $Sp(N)$ CFT$_3$.
\end{abstract}
\vspace{0.5cm}

\vspace{1.0cm}

\end{center}

\end{titlepage}

\pagestyle{empty}
\pagestyle{plain}

\def\vx{{\vec x}}
\def\ip{${\cal I}^+$}
\def\p{\partial}
\def\po{$\cal P_O$}

\setcounter{page}{1}
\pagenumbering{arabic}
\tableofcontents
\section{Introduction}
The conjectured dS/CFT correspondence attempts to adapt the wonderful successes of the AdS/CFT correspondence to universes (possibly like our own) which exponentially expand in the far future.  The hope \cite{Strominger:2001pn, Hull:1998vg,Witten:2001kn,Maldacena:2002vr,Harlow:2011ke} is to define  bulk de Sitter (dS) quantum gravity in terms of a holographically dual CFT living at \ip\ of dS, which is the asymptotic conformal boundary at future null infinity.  A major obstacle to this program has been the absence of any explicit microscopic realization.  This has so far  prevented the detailed development of the dS/CFT dictionary.  This situation has recently been improved by an
explicit  proposal \cite{Anninos:2011ui} relating Vasiliev's higher-spin gravity in dS$_4$ \cite{Vasiliev:1990en,Vasiliev:1999ba} to the dual $Sp(N)$ CFT$_3$ described in \cite{LeClair:2007iy}.
In this paper we will use this higher-spin context to write some new entries in the dS/CFT dictionary.

The recent proposal \cite{Anninos:2011ui} for a microscopic realization of dS/CFT begins with the duality relating the
free (critical) $O(N)$ CFT$_3$ to higher-spin gravity on AdS$_4$ with Neumann (Dirichlet) boundary conditions on the scalar field.  Higher-spin gravity - unlike string theory \cite{Hull:1998vg} -  has a simple analytic continuation from negative to positive cosmological constant $\Lambda$.  Under this continuation, AdS$_4 \to$ dS$_4$ and the (singlet) boundary CFT$_3$ correlators are simply transformed by the replacement of $N\to -N$.  These same transformed correlators arise from the $Sp(N)$ models constructed from anticommuting scalars. It follows that the
free (critical) $Sp(N)$ correlators equal those of higher-spin gravity on dS$_4$ with future Neumann (Dirichlet) scalar boundary conditions (of the type described in \cite{Anninos:2011jp}) at \ip.

This mathematical relation between the bulk dS and boundary $Sp(N)$ correlators may provide a good starting point for understanding quantum gravity on dS, but so  far important physical questions remain unanswered. For example we do not know how to relate these physically $un$measurable correlators to a set of true physical observables or to the dS horizon entropy. These crucial entries in the dS/CFT dictionary are yet to be written.

As a step in this direction,  in this paper we investigate the relation between quantum states in the bulk higher-spin gravity and those in the boundary CFT$_3$.   Bulk higher-spin gravity has fields of $\Phi^s$ with all even spins
 $s=0,2,....$, which are dual to CFT$_3$ operators $\co^s$ with the same spins.
In the CFT$_3$, we can also associate a state to each operator by the state-operator correspondence. One way to do this is to take the southern hemisphere of $S^3$, insert the operator $\co^s$ at the south pole, and then define a state $\Psi^s_{S^2}$ as a functional of the boundary conditions on the equatorial $S^2$. For every object in the CFT$_3$, we expect a holographically dual object in the bulk dS$_4$ theory. This raises the question: what is the bulk representation of the spin-$s$ state $\Psi^s_{S^2}$?

In Lorentzian AdS$_4$  holography, the state created by a primary operator ${\cal O}$ in the CFT$_{3}$ on S$^{2}$ has, at weak coupling, a bulk representation as the single particle state of the field $\Phi$ dual to $\co$ with a smooth minimal-energy wavefunction localized near the center of AdS$_4$.  The form of the wavefunction is dictated by the conformal symmetry.

In dS$_4$ holography, the situation is rather different.
States in dS$_4$ quantum gravity are usually thought of as wavefunctions on complete spacelike slices which are topologically $S^3$.\footnote{As explained in \cite{Maldacena:2002vr,Harlow:2011ke}, such states do play an important role in dS/CFT, but as generating functions for correlators rather than as duals to CFT$_3$ states on $S^2$. The relation between the $R^3$ and $S^3$  bulk states in our example is detailed below.}   These do not seem to be good candidates for bulk duals to $\Psi^s_{S^2}$ because, among other reasons,  they are not associated to any $S^2$ in \ip. However, dS$_4$ also has everywhere spacelike and  geodesically complete $R^3$  slices which end at  an $S^2$ in \ip. Here we propose a construction of  the bulk version of  $\Psi^s_{S^2}$ on these slices, again as single particle states whose form is dictated by the conformal symmetry.  Interestingly, the classical wavefunction for the particle turns out to be the (anti) quasinormal modes for the static patch of de Sitter, as constructed in \cite{LopezOrtega:2006my,Anninos:2011af}.\footnote{We are grateful to D. Anninos for pointing this out \cite{dio}.}

This relation between bulk and boundary states has a potentially profound nonperturbative consequence briefly mentioned  in section 4.1 \cite{mkss}.  The operator $\co^0$ dual to a scalar $\Phi$ is bilinear in boundary fermions and hence obeys $(\co^0)^{{N \over 2}+1}=0$. Under bulk-boundary duality this translates into an $N \over 2$-adicity relation for $\Phi$: one cannot put more than
$N \over 2$ bulk scalar quanta into the associated quasinormal mode. Further investigation of this dS exclusion principle is deferred to later work.

We also construct a norm for these bulk states and show that it is the Zamolodchikov norm on $S^3$ of the CFT$_3$ operator $\co^s$.  Explicit formulae are exhibited only for the scalar $s=0$ case but we expect the construction to generalize to all $s$.

This paper is organized as follows. In section 2 we revisit the issue of the usual global dS-invariant vacua for a free massive scalar field,
paying particular attention to the case of $m^2\ell^2=2$ (where $\ell$ is the de Sitter radius) arising in higher spin gravity. The invariant vacua include the familiar Bunch-Davies Euclidean vacuum $|0_E\>$, as well as a pair of $|0^\pm\>$ of in/out  vacua with no particle production. As the scalar field  acting on $|0^-\>$ ($|0^+\>$) obeys Dirichlet (Neumann) boundary conditions on \ip, these are related to the critical (free) $Sp(N)$ model. Generically all dS-invariant vacua are Bogolyubov transformations of one another, but we find that at $m^2\ell^2=2$ the transformation is singular and the in/out vacua are non-normalizable plane-wave type states. In section 3 we use the conformal symmetries to find the classical bulk wavefunctions associated to an operator insertion on \ip, and note the relation to (anti) quasinormal modes. The construction uses a rescaled bulk-to-boundary Green function defined with Neumann or Dirichlet \ip\ boundary conditions. We also show that the Klein-Gordon inner product of these wavefunctions agrees with the conformally-covariant CFT$_3$ operator two-point function.  In section 4 we consider the Hilbert space on $R^3$ slices ending on an $S^2$ on \ip. This Hilbert space was explicitly constructed in \cite{Sasaki:1994yt} for a free scalar on hyperbolic slices ending on \ip. There are two such  Hilbert spaces, which we denote the northern and southern Hilbert space, which live on spatial $R^3$ slices extending to the  north or south of the $S^2$. The northern and southern slices add up to a global $S^3$. Hence the tensor product of the northern and southern Hilbert spaces is  the global Hilbert space on $S^3$, much as the left and right Rindler Hilbert spaces tensor to the global Minkowski Hilbert space. We show that the global  $|0^\pm\>$ vacua are simple tensor products of the northern and southern Dirichlet and Neumann vacua. We then use symmetries to uniquely identify the states of the southern Hilbert space with those of the free and critical $Sp(N)$ models on an $S^2$. This leads directly to the dS exclusion principle. We further construct an inner product for the southern Hilbert space which agrees, for states dual to \ip\ operator insertions, to the conformal two-point function on \ip. In section 5 we discuss the restriction of Euclidean vacuum to a southern state and recall from \cite{Sasaki:1994yt}, that this is a mixed state which is thermal with respect to an $SO(3,1)$ Casimir.  It would be interesting to relate this result to dS entropy in the present context.  In section 6 we show that the standard CFT$_3$ state-operator correspondence maps the known pseudo-unitary C-norm of the $Sp(N)$ model to the Zamolodchikov two-point function. This completes the demonstration that the bulk states on $R^3_S$ have the requisite properties to be dual to the boundary $Sp(N)$ CFT$_3$ states on $S^2$.  Speculations are made on the possible relevance of pseudo-unitarity to the consistency of dS/CFT in general. An appendix gives some explicit formulae for the $SO(4,1)$ Killing vectors of dS$_4$.

\section{Global dS vacua at $m^2\ell^2=2$}
    In this section we describe the quantum  theory of a free scalar field $\Phi$ in dS$_4$ with wave equation
  \be (\nabla^2-m^2)\Phi=0, \ee
and     mass
\be m^2\ell^2=2. \ee This is the case of interest for Vasiliev's higher-spin gravity. While there have been many general discussions of this problem, peculiar singular behavior as well as simplifications appear at
the critical value $m^2\ell^2=2$  which are highly relevant to the structure of dS/CFT. A parallel discussion of de Sitter vacua and scalar Green functions in the context of dS/CFT was given in \cite{Bousso:2001mw}. However that paper in many places specialized to the large mass regime  $m^2\ell^2>{9 \over 4}$ , excluding the region of current interest.
The behavior in the region $m^2\ell^2<{9 \over 4}$ divides into  three cases $m^2\ell^2>{2}$, $m^2\ell^2=2$ and $m^2\ell^2<2$.
Much of the structure we describe below pertains to the entire range $m^2\ell^2<{9 \over 4}$ with an additional  branch-cut prescription for the Green functions.

\subsection{Modes}
  We will work  in the dS$_4$ global coordinates
\be \label{gb} {ds^2 \over \ell^2}=-dt^2+\cosh^2 t d^2\Omega_3
=-dt^2+\cosh^2 t \left[d\psi^2+\sin^2{\psi} \left( d\theta^2+\sin^2{\theta} d\phi^2\right) \right]
, \ee where ${\O}^i \sim (\psi,\theta,\phi)$ are coordinates on the global $S^3$ slices.  Following the notation of   \cite{Bousso:2001mw} solutions of the wave equation
can be expanded in modes  \be \phi_{Lj}(x)=y_L(t)Y_{Lj}(\Omega) \ee of total  angular momentum $L$ and spin labeled by the multi-index $j$. The spherical harmonics $Y_{Lj}$ obey
\bea Y^*_{Lj}(\O)&=&(-)^LY_{Lj}(\O)=Y_{Lj}(\O_A),\cr
        D^2Y_{Lj}(\O)&=&-L(L+2)Y_{Lj}(\O),\cr
        \int_{S^3}\sqrt{h}d^3\Omega Y^*_{Lj}(\O)Y_{L'j'}(\O)&=&\delta_{L,L'}\delta_{j,j'},\cr
          \sum Y^*_{Lj}(\O)Y_{Lj}(\O')&=&{1\over {\sqrt h}}\delta^3(\O-\O'),
  \eea
  where $\sqrt{h}$ and $D^2$ are the measure and Laplacian on the unit $S^3$, $\O_A$ is the antipodal point of $\Omega$, and here and hereafter $\sum$ denotes summation over all allowed values of $L$ and $j$.
The time dependence is then governed by the second order ODE
\be\label{hio} \p_t^2y_L+3\tanh t \p_t y_L+\left(m^2\ell^2+{L(L+2) \over\cosh^2t}\right)y_L=0.\ee
\subsubsection{Neumann and Dirichlet modes  }
Eq.~(\ref{hio}) has the $real$ solutions
\be y^\pm_L={2^{L+h_\pm+\half}}(L+1)^{\pm \half}\cosh^L te^{-(L+h_\pm) t}F(L+{3\over2},L+h_\pm, h_\pm -\frac{1}{2};-e^{-2t}) \ee
where
\be
h_{\pm} \equiv \frac{3}{2} \pm  \sqrt{\frac{9}{4}-m^2 \ell^2}.
\ee
We are interested in $m^2\ell^2=2$, which implies
\be h_-=1,~~~~h_+=2,\ee and
\be
y^\pm_L =  \frac{ (-i)^{\half \pm \half} 2^L}{\sqrt{1+L}}
 \cosh^L t e^{-(L+1)t}
 \left[
\frac{1}{\left(1-i e^{-t}\right)^{2L+2}}
 \mp \frac{1}{\left(1+i e^{-t}\right)^{2L+2}}
 \right].
\ee
The modes  behave near \ip\ as $e^{-h_\pm t}$
\bea \label{asm} t&\to& \infty,~~~~~\cr y_L^-&\to&(2(L+1)^{- \half} )e^{-t}+\co(e^{-3t})~~~~~{\rm Neumann},\cr~~~~~y_L^+&\to& (4(L+1)^{ \half} )e^{-2t}+\co(e^{-4t})~~~~~{\rm Dirichlet}.\eea
Accordingly we refer to the $+$ modes as Dirichlet and the $-$ modes as Neumann.
We have normalized so that the Klein-Gordon inner product is  \be\label{nm} \langle\phi^+_{Lj} |\phi^-_{L'j'}\rangle_{S^3}\equiv i \int_{S^3}d^3\Sigma^\mu \phi^{+*}_{Lj}\overleftrightarrow{ \p_\mu }\phi^-_{L'j'}   = i \delta_{LL'}\delta_{jj'},\ee
with $d^3\Sigma^\mu$ the induced measure times the normal to the $S^3$ slice.

Under time reversal
\be \label{df}y_L^\pm(t)=\pm (-)^Ly_L^\pm(-t),\ee
so that
\be \label{nm2}\phi^\pm_{Lj}(x)=\pm\phi^\pm_{Lj}(x_A)=(-)^L\phi^{\pm*}_{Lj}(x),\ee
where the point $x_A$ is antipodal to the point $x$.
This implies that an incoming Dirichlet (Neumann) mode propagates to an outgoing
Dirichlet (Neumann) mode. This is not the case for generic $m^2$ and, as will be seen below, allows for Dirichlet and Neumann vacua with no particle production.

\subsubsection{Euclidean modes }
  Euclidean modes are defined by the condition that when dS$_4$ is analytically continued to $S^4$ they remain nonsingular on the southern hemisphere. In other words
\be y^E_L(t=-{i\pi \over 2})={\rm nonsingular}.\ee
One finds that the combination
%\be y^E_L=\frac{y^-_L+iy^+_L}{\sqrt{2}}=\frac{2^{L+1}}{\sqrt{2L+2}}{\Gamma(L+2)\Gamma(L+{3 \over 2})\over \Gamma(2L+3)\Gamma(\half)}\cosh^L te^{-(L+1)t}F(L+{3\over2},L+1,  2L+3;1+e^{-2t}) \ee
%\be y^E_L=\frac{y^-_L+iy^+_L}{\sqrt{2}}=\frac{2^{-\frac{3}{2}-L}}{\sqrt{1+L}}\cosh^L te^{-(L+1)t}F(L+{3\over2},L+1,  2L+3;1+e^{-2t}) \ee
\be y^E_L
=\frac{y^-_L+iy^+_L}{\sqrt{2}}
%&=&\frac{2^{-\frac{3}{2}-L}}{\sqrt{1+L}}\cosh^L te^{-(L+1)t}F(L+{3\over2},L+1,  2L+3;1+e^{-2t}) \nonumber\\
=\frac{2^{L+1}}{\sqrt{2L+2}}\frac{\cosh^L t e^{-(L+1)t}}{\left(1-i e^{-t}\right)^{2L+2}}
\ee
is nonsingular at $t=-i\pi/2$.
Hence $y_L^\mp$ are simply the real and imaginary parts of the $y^E_L$. (\ref{df}) and (\ref{nm2})
imply the relations
\be y^{E*}_L(t)=(-)^{L+1}y^E_L(-t)\ee
\be \phi^{E}_{Lj}(x_A)=(-)^{L+1}\phi^{E*}_{Lj}(x).\ee
\be \langle \phi^E_{Lj}|\phi^E_{L'j'}\rangle_{S^3}=\delta_{LL'}\delta_{jj'}\ee
\subsection{Vacua}
In the quantum theory $\Phi$ is promoted to an operator which we denote $\hat \Phi$
obeying the equal time commutation relation
\be [ \hat \Phi(\O,t),\p_t\hat \Phi(\O',t)]={i \over  \sqrt{h}\cosh^3 t}\delta^3(\Omega-\O').\ee
Defining annihilation and creation operators
\be a^E_{Lj}=\langle \phi^E_{Lj}|\hat \Phi\rangle_{S^3},~~~~ a^{E\dagger}_{Lj}=-\langle \phi^{E*}_{Lj}|\hat \Phi \rangle_{S^3},\ee
the global Euclidean (or Bunch-Davies) vacuum is defined by
\be a^E_{Lj}|0_E\rangle =0. \ee
We normalize so that $\langle 0_E |0_E\rangle=1$.
For any $m^2$ there is a  family of  dS-invariant vacua labeled by a complex parameter
$\alpha$.  They are annihilated by the normalized Bogolyubov-transformed oscillators
\be a^\alpha_{Lj}={1 \over \sqrt {1-e^{\alpha+\alpha^*}}}\left( a^{E}_{Lj} -e^{\alpha^*} a^{E\dagger}_{Lj}\right). \ee
We are interested in the vacua annihilated by the Dirichlet or Neumann modes for the case of $m^2 \ell^2=2$, which correspond to $e^\alpha=\pm 1$. In that case the Bogolyubov transformation is singular. Nevertheless we can still construct non-normalizable plane-wave type vacua as follows.

The field operator may be decomposed as
\be \hat \Phi=\hat \Phi^++\hat \Phi^-,\ee
where $\hat \Phi^\pm \sim e^{-h_\pm t}$ near \ip. The squeezed states
\be |0^\pm \rangle = e^{\pm \half \sum (-)^L (a^{E\dagger}_{Lj})^2}|0_E\rangle
\ee
%\be |0^\pm \rangle =N_0 e^{\pm \half \sum (-)^L (a^{E\dagger}_{Lj})^2}|0_E\rangle
%,\quad N_0\equiv\prod_L 2^{\frac{1}{4}}
%\ee
then obey  \be\label{bxz} \hat \Phi ^-|0^-\rangle =0~~~~~~{\rm Dirichlet}, \ee
 \be\label{bxdz} \hat \Phi ^+|0^+\rangle =0~~~~~~{\rm Neumann}. \ee
Since only Dirichlet (Neumann) modes act non-trivially on $|0^-\rangle$ ($ |0^+ \rangle$) we refer to it as the Dirichlet (Neumann) vacuum. These vacua are dS invariant. With the conventional norm, $\hat \Phi^\pm$  are hermitian and their eigenstates are non-normalizable. Generalized dS non-invariant plane-wave type Neumann states with nonzero eigenvalues for  $\hat\Phi^+$
\be \hat \Phi^+|\Phi^+\rangle = \Phi^+ | \Phi^+\rangle \ee
are constructed as
\be |\Phi^+\rangle =e^{-\langle \Phi^+|\hat \Phi^-\rangle_{S^3}} |0^+\rangle.\ee $\Phi^+$ here is an arbitrary solution of the classical wave equation,  which can be parameterized by an arbitrary function $\Phi^+(\O)$ on \ip
\be t\to \infty,~~~\Phi^+(\O,t)\to \Phi^+(\O)e^{-h_+t}.\ee %We note that the Klein-Gordon inner product for classical wavefunctions
The states are delta-functional normalizable with respect the usual inner product
\be \langle \Phi^+|{\Phi^{+}}'\rangle=\delta\left( \Phi^+-{\Phi^+}' \right),\ee
where the delta function integrates to one with the measure
\be {\cal D}\Phi^+ \equiv \prod_{L,j}\frac{dc^+_{Lj}}{\sqrt{\pi}}, ~~~\Phi^+(x)=\sum c^+_{Lj}\phi^+_{Lj}(x).\ee The $c^+_{Lj}$ satisfies the reality condition $c^{+*}_{Lj}=c^+_{Lj}(-)^L$.
%\be {\cal D}\Phi^+ \equiv \prod_{L,j}\frac{dc^+_{Lj}}{\sqrt{2\pi}}. ~~~\Phi^+(x)=\sum c^+_{Lj}\phi^+_{Lj}(x).\ee
One may similarly  define generalized  Dirichlet states obeying
\be \hat \Phi^-|\Phi^-\rangle = \Phi^- | \Phi^-\rangle.\ee

The Euclidean vacuum can be expressed in terms of $|0^\pm \rangle$ as
\be |0_E>%=N_0 e^{ -{1\over 2}\sum (-)^L a^\pm_{Lj}a^\pm_{Lj}}|0^\mp  \rangle
 =\int {\cal D}\Phi^\pm
e^{\mp {1\over 16} \int d^3\O d^3\O'\Phi^\pm(\O)\Delta_\mp(\O,\O')\Phi^\pm(\O')}|\Phi^\pm\rangle,\ee
where\bea
%a^{\pm}_{Lj}&\equiv & \pm i \langle \phi^{\mp}_{Lj}|\hat \Phi\rangle_{S^3} ,\quad
\Delta_{\pm} (\O, \O')&=&\mp \sum Y^*_{Lj}(\O)Y_{Lj} (\O') (2L+2)^{\pm 1}
=\frac{1}{2^{2\mp1}\pi^2}\frac{1}{\left(1-\cos{\Theta_3}\right)^{h_\pm}}, \nonumber\\
\cos{\Theta_3(\Omega,\Omega')} &\equiv &
\cos{\psi}\cos{\psi'} + \sin{\psi}\sin{\psi'}(\cos{\theta}\cos{\theta'} + \sin{\theta}\sin{\theta'}\cos{(\phi-\phi')}).\nonumber\\
\eea $\Delta_{\pm} $ are the (everywhere positive) two-point functions for a CFT$_3$ operator with $h_+=2$ and $h_-=1$.\footnote{Here, we regulate the expressions of $\Delta_{\pm}$ as sums over the spherical harmonics by introducing $e^{-L \epsilon}$ in each term in the sum and take the limit of $\epsilon \rightarrow 0$ at the end after the summation.
}
These satisfy
\be
-\int \sqrt{h}d^3\Omega'' \Delta_+(\Omega,\Omega'')\Delta_-(\Omega'',\Omega')=\frac{1}{\sqrt{h}}\delta^3(\Omega-\Omega').
\ee
We also have the relations
\be |\Phi^+\rangle = \frac{1}{N_0}\int {\cal D}\Phi^-e^{\langle \Phi^-|\Phi^+\rangle_{S^3}} |\Phi^-\rangle,\quad
N_0 \equiv \prod_{L,j} \sqrt{2},
\ee
%\be |\Phi^+\rangle = \int {\cal D}\Phi^-e^{\langle \Phi^-|\Phi^+\rangle_{S^3}} |\Phi^-\rangle,\ee
\be \langle \Phi^- |\Phi^+\rangle_{S^3} = \frac{1}{N_0}e^{\langle \Phi^-|\Phi^+\rangle_{S^3}} .\ee
In particular
%\be \langle0^- |0^+\rangle_{S^3} =1.\ee
\be \langle0^- |0^+\rangle_{S^3} =\frac{1}{N_0}.\ee

The Wightman function in the Euclidean vacuum is
\bea \label{ge}  G_E(x;x')&=&\sum \phi^{E}_{Lj}(x)\phi^{E*}_{Lj}(x')\cr &=&\sum (-)^{L+1}\phi^{E}_{Lj}(x)\phi^E_{Lj}(x'_A)\cr &=&\half\sum (-)^{L} \left( \phi^{-}_{Lj}(x)\phi^-_{Lj}(x')+\phi^{+}_{Lj}(x)\phi^+_{Lj}(x')+i\phi^{+}_{Lj}(x)\phi^-_{Lj}(x')-i\phi^{-}_{Lj}(x)\phi^+_{Lj}(x')\right).\cr && \eea
In terms of the dS-invariant distance function
\be
P(t,\Omega;t',\Omega') = \cosh{t} \cosh{t'} \cos\Theta_3(\Omega,\Omega')
-\sinh{t} \sinh{t'},
\ee %where
%\bea
%\cos{\Theta_3(\Omega,\Omega')} &\equiv &
%\cos{\psi}\cos{\psi'} + \sin{\psi}\sin{\psi'}\cos{\theta}(\cos{\theta'} + \sin{\theta}\sin{\theta'}\cos{(\phi-\phi')}).
%\eea
this becomes simply
\be \label{ge2}G_E(x;x')= \frac{1}{8\pi^2} \frac{1}{1-P(x;x')},\ee
with the usual $i \epsilon$ prescription for the singularity.

\section{Boundary operators and quasinormal modes}
According to the dS$_4$/CFT$_3$ dictionary, for every spin zero primary CFT$_3$ operator $\co$ of conformal weight $h$ there is a bulk scalar field $\Phi$ with mass
$m^2\ell^2=h(3-h)$. Boundary correlators of $\co$ are then related by a  rescaling to bulk $\Phi$ correlators whose arguments are pushed to the boundary at \ip.  As in AdS/CFT, a particular classical bulk wavefunction of $\Phi$ can be associated to a boundary insertion of $\co$ (at the linearized level) by symmetries: it must scale with weight $h$ under the isometry corresponding to dilations, obey the lowest-weight condition, and  be invariant under rotations around the point of the boundary insertion. The resulting  wavefunction is a type of bulk-to-boundary Green function. Interestingly\cite{dio}, the (lowest) highest-weight modes can also be identified as (anti) quasinormal modes for the static patch of de Sitter, as constructed in \cite{LopezOrtega:2006my,Anninos:2011af}. In this section we  determine this wavefunction explicitly, regulate the singularities, generalize it to multi-particle insertions and define a symplectic product. In the following section we will then use these classical objects to construct the associated  dual bulk quantum states and their inner products.

\subsection{Highest and lowest weight wavefunctions}

In this subsection we give expressions for the classical wavefunctions, associated to lowest (highest) weight primary operator insertions at the south (north) pole of \ip\ in terms of
rescaled Green functions in the limit that one argument is pushed to \ip.  These wavefunctions each comes in a Neumann and a Dirichlet flavor, denoted  $\Phi^\pm_{lw}(x)$ ($\Phi^\pm_{hw}(x)$) depending on whether the weight of the dual operator insertion is $h_+$ or $h_-$.

The  relevant Green functions are\footnote{
Note that $G_+ (G_-)$ is even (odd) under the antipodal map. Combinations of the Euclidean
Green function with such properties have been previously studied in the context of elliptic $Z_2$-identification of de Sitter space \cite{Sanchez:1986gr,Gibbons:1986dd,Parikh:2002py} which may be related to our construction.
}
  \be G_\pm (x;x')\equiv G_E(x;x')\pm G_E(x;x'_A) =\frac{1}{8\pi^2}\left( \frac{1}{1-P(x;x')}\pm \frac{1}{1+P(x;x')}\right) \ee
with $G_E$ the Wightman function for the  Euclidean  vacuum given in equation (\ref{ge2}). $G_-$ ($G_+$) obeys Neumann (Dirichlet) boundary conditions at \ip\ away from $x=x'$.
These are for $m^2\ell^2=2$ the Green functions with future boundary conditions as discussed in \cite{Anninos:2011jp}. We have normalized them so that they have the Hadamard form for the short-distance singularity. In the Neumann case we begin with $G_-$, which is (using the mode decomposition (\ref{ge})) given by
\be\label{eqGminus}
 G_-(x;x')=\sum (-)^{L} \left( \phi^{-}_{Lj}(x)\phi^-_{Lj}(x')+i\phi^{+}_{Lj}(x)\phi^-_{Lj}(x')\right).\ee
From this we construct the rescaled Green function
\be \Phi^-_{lw}(x;t')=e^{h_-t'}G_-(x;t',\O_{SP}),\ee
in which the second argument is placed at the south pole $\O_{SP}$ where $\psi'=0$.
One may then check that (ignoring singularity prescriptions)
 \be \label{lwd}\Phi^-_{lw}(x)\equiv \lim_{t'\to \infty} \Phi^-_{lw}(x;t')= \frac{1}{2\pi^2(\sinh t-\cos \psi \cosh t)}.\ee
%the mode sum in the last line of (\ref{ge})
Using (\ref{eqGminus}) and the asymptotics (\ref{asm}) one finds that  near \ip (not ignoring singularities)
\bea \Phi^-_{lw}(x)&=&8\sum (-)^{L} \left( {e^{-t} \over 2L+2} Y_{Lj}(\O)Y_{Lj}(\O_{SP})+ie^{-2t}Y_{Lj}(\O)Y_{Lj}(\O_{SP})+\co(e^{-3t})\right)\cr &=&8e^{-t}\Delta_-(\O,\O_{SP})+8i {e^{-2t}\over \sqrt{h}}\delta^3(\O-\O_{SP})+\co(e^{-3t})\eea

Let us now confirm that $\Phi^-_{lw}(x)$ has the same symmetries as an insertion of a primary
operator $\co(\O_{SP})$ at the south pole of \ip. First we note that the choice of a point on \ip\ breaks
$SO(4,1)$ to $SO(3)\times SO(1,1)$. Both $\Phi^-_{lw}(x)$ and $\co(\O_{SP})$ are manifestly invariant under the $SO(3)$ spatial rotations.  The generator of $SO(1,1)$ dilations, denoted $L_0$, acts on $\co(\O_{SP})$ as
\be [L_0,\co(\O_{SP})]=h_-\co(\O_{SP}).\ee
In the bulk it is generated by the Killing vector field
%\be L_0=\tanh t \sin \psi \p_\psi -\cos \psi \p_t,\ee
\be L_0=\cos \psi \p_t-\tanh t \sin \psi \p_\psi ,\ee
where the south pole is $\psi=0$.  dS invariance implies
\be (L_0-L_0')G_-(x,x')=0.\ee
It follows from this together with the definition (\ref{lwd}) that the  wavefunction obeys
\be L_0\Phi^-_{lw}(x)=h_-\Phi^-_{lw}(x). \ee
By construction it obeys the wave equation
\be (\nabla^2-m^2)\Phi^-_{lw}(x)=0. \ee
Acting on $SO(3)$ invariant symmetric functions we have
\be \ell^2\nabla^2=-L_0(L_0-3) +M_{-k}M_{+k},\ee
where the 6 Killing vector fields $M_{\pm k}$ (given in Appendix \ref{appendix1}) are the raising and lowering operators for $L_0$ and we sum over $k$.
It then follows that
\be M_{-k}M_{+k}\Phi^-_{lw}(x)=\left( m^2\ell^2-h_-(3-h_-)\right)\Phi^-_{lw}(x)=0,\ee
and hence \be
M_{+k}\Phi^-_{lw}(x)=0.\ee
which corresponds to the lowest-weight condition for the $\co$
\be [M_{+k},\co(\O_{SP})]=0.\ee
It may be shown that these symmetries uniquely determine the solution. Hence $\Phi^-_{lw}$ is identified as the classical wavefunction associated to the insertion of the primary $\co$ at the south pole.

A parallel argument leads to the dual of a highest weight operator insertion at the north pole
. The wavefunction is
\be \Phi^-_{hw}(x;t')=\lim_{t'\to \infty} e^{h_-t'}G_-(x;t',\O_{NP}).\ee
This obeys the relations
\be
M_{-k}\Phi^-_{hw}(x)=0,~~~~ L_0\Phi^-_{hw}(x)=-h_-\Phi^-_{hw}(x), \ee
and has the asymptotic behavior
\be \Phi^-_{hw}(x)=8e^{-t}\Delta_-(\O,\O_{NP})+8i {e^{-2t}\over \sqrt{h}}\delta^3(\O-\O_{NP})+\co(e^{-3t}).\ee
%\be \Phi^-_{hw}(x)=4e^{-2t}\Delta_+(\O,\O_{NP})+4i {e^{-t}\over \sqrt{h}}\delta^3(\O-\O_{NP})+\co(e^{-3t})\ee

Similar formulae apply to the Dirichlet case by beginning with  $G_+$ in the above construction and replacing $+\leftrightarrow -$. For example
\be \Phi^+_{hw}(x)=-8e^{-2t}\Delta_+(\O,\O_{NP})-8i {e^{-t}\over \sqrt{h}}\delta^3(\O-\O_{NP})+\co(e^{-3t}).\ee

We see from the above that the highest-weight wavefunction is  smooth on the future horizon of the southern static patch dS$_4$, and hence related to the quasinormal modes found in \cite{LopezOrtega:2006my,Anninos:2011af}. The lowest quasinormal mode which is invariant under the $SO(3)$ of the static dS$_4$ is exactly the $\Phi_{hw}^-$ with $h_-=1$ while the second lowest $SO(3)$-invariant quasinormal mode corresponds to $\Phi_{hw}^+$ with $h_+=2$.
Lowest weight states are smooth on the past horizon and hence related to anti-quasinormal modes.
%\footnote{
%Similar relation between the lowest quasinormal modes and (some notion of) highest-weight states in the BTZ black hole was observed in \cite{Sachs:2008gt}.
%}

\subsection{General multi-operator insertions}

In the preceding subsections we found the bulk duals of primary operators inserted at the north/south pole in the coordinates (\ref{gb}). This can be generalized to insertions at an arbitrary  point  on \ip\ with a general time slicing near \ip. Let us introduce coordinates  $x\sim({y}^i,t)$ such that near \ip
\be \label{indm} ds_4^2 \to -dt^2+e^{2t}h_{ij}(y)d{y}^id{y}^j,~~i,j=1,2,3.\ee
The dual wavefunction is then the $t'\to \infty$ limit of the rescaled Green function, denoted by
\be \Phi^\pm_{y_1}(x)=\lim_{t'\to \infty} e^{h_\pm t'}G_\pm(x;t',y_1).\ee
For the special cases of operator insertions at the north or south pole in global coordinates  these reduce  to our previous expressions.
Note that coordinate transformations of the form $t\to t+f(y)$ induce a conformal transformation on \ip
\be \label{cnf} h_{ij} \to e^{2f(y)} h_{ij}, ~~~~ \Phi^\pm_{y_1}(x)\to e^{h_\pm f(y)} \Phi_{y_1}(x), \ee  as appropriate for a conformal field of weight $h_\pm$. Hence the relative normalization in (\ref{lwd}) will depend on the conformal frame at \ip.

One may also consider multi-operator insertions such as $\co(y_1)\co(y_2)$ in the CFT$_3$ at \ip.  At the level of free field theory considered here these are associated  to a bilocal wavefunction in the product of two bulk scalar fields
\be \Phi_{ y_1}(x_1)\Phi_{y_2} (x_2).\ee

We will use $\Phi_\O$ to denote these wavefunctions when working in global coordinates (\ref{gb}).
We note that in such coordinates near \ip\ for an insertion at a general point \be \label{fc} \Phi^\pm_{\O_1}(t,\O)=\mp 8e^{-h_\pm t}\Delta_\pm(\O,\O_{1})\mp 8i {e^{-h_\mp t}\over \sqrt{h}}\delta^3(\O-\O_{1})+\co(e^{-3t}).\ee

\subsection{Klein-Gordon inner product}
We wish to define an inner product between e.g. two Neumann wavefunctions $\Phi^-_{{\O}_1}$ and $\Phi^-_{{\O}_2} $. Later on we will compare this to the  inner product on the CFT$_3$ Hilbert space and the two-point function of $\co$ on $S^3$.
One choice is to take a global spacelike $S^3$ slice in the interior and define
the Klein-Gordon inner product
\be \left\langle\Phi^-_{{\O}_1}|\Phi^-_{{\O}_2}\right\rangle_{S^3}\equiv i \int_{S^3}d^3\Sigma^\mu \Phi^{-*}_{{\O}_1}\overleftrightarrow{ \p_\mu }\Phi^-_{{\O}_2} .\ee
This integral does not depend on the choice of $S^3$ which can be pushed up to \ip. One may then see immediately from (\ref{fc}) that there are two nonzero terms proportional to $\Delta_-$ giving
\be \left\langle\Phi^-_{{\O}_1}|\Phi^-_{{\O}_2}\right\rangle_{S^3}= 16 \Delta_-(\O_1-\O_2) \ee

One may also  define an inner product not on global spacelike $S^3$ slices, but on a spacelike $R^3$ slice which ends on an $S^2$ on \ip.   The result is invariant under any deformation of the $S^2$ which does not cross the insertion point. To be definite, we take the $S^2$ to be the equator, ${\O}_1$ to be  in the northern  hemisphere and ${\O}_2$ to be  in the southern hemisphere, and the slice to be $R^3_S$ which intersects the south pole.
One then finds, pushing $R_S^3$ up to the southern hemisphere of \ip
\be \left\langle\Phi^-_{{\O}_1}|\Phi^-_{{\O}_2}\right\rangle_{R_S^3}\equiv i \int_{R_S^3}d^3\Sigma^\mu \Phi^{-*}_{{\O}_1}\overleftrightarrow{ \p_\mu }\Phi^-_{{\O}_2}= 8 \Delta_-(\O_1-\O_2) .\ee

Similarly, the inner product between two Dirichlet wavefunctions is given by
\be \left\langle\Phi^+_{{\O}_1}|\Phi^+_{{\O}_2}\right\rangle_{R_S^3}=- 8 \Delta_+(\O_1-\O_2) .\ee

 \section{The southern Hilbert space  }

 We now turn to the issue of bulk quantum states.  Quantum states in dS are often discussed, as in section 2,  in terms of a Hilbert space built on the global $S^3$ slices. The structure of the vacua and Green functions for such states was described in section 2.   However dS has the unusual feature that there are geodesically complete topologically $R^3$ spacelike slices which end on an $S^2$ in \ip, which we will typically take to be the equator.  Examples of these are the hyperbolic slices, the quantization on which was studied in detail in \cite{Sasaki:1994yt}. We will see that the quantum states built on these $R^3$ slices are  natural objects in dS/CFT.
An $S^2$ in \ip\ is in general the boundary of  a ``northern" slice, denoted $R^3_N$ and a ``southern"  slice denoted $R^3_S$. The topological sum obeys $R^3_S \cup R^3_N=S^3$. Hence the relation of the southern and northern Hilbert spaces on $R^3_S$ and $R^3_N$ to that on $S^3$ is like that of the left and right Rindler wedges to that of global Minkowski space. It is also like the relation of the Hilbert spaces of the northern and southern causal diamonds to that of global dS. However the diamond Hilbert spaces in dS quantum gravity are problematic  in quantum gravity with a fluctuating metric because it is hard to find sensible boundary conditions.

A strong motivation for considering the $R^3_{S,N}$ slices comes from the picture of a state in the boundary CFT$_3$. The state-operator correspondence in CFT$_3$ begins with an insertion of a (primary or descendant)  operator $\co$ at the south pole of $S^3$, and then defines a quantum state as a functional of the boundary conditions on an $S^2$ surrounding the south pole. For every object in the CFT$_3$, we expect a holographically dual object in the bulk dS$_4$ theory.  The dual bulk quantum state must somehow depend on the choice of $S^2$ in \ip.  Hence it is natural to define the bulk state on the $R^3$ slice which ends on this  $S^2$ in \ip.  This is how holography works in AdS/CFT:
CFT states live on the boundaries of the spacelike slices used to define the bulk states.
\subsection{States}
In order to define quantum states on $R^3_S$, we first note that modes of the scalar field operator
$\hat \Phi(\O,t)$ are labeled by operators $\hat\Phi^\pm( \O )$ defined on \ip\ via the relation
\be \label{ipo} \lim_{t\to \infty}\hat \Phi(\O,t)=e^{-h_+t}\hat \Phi^+(\O)+e^{-h_-t}\hat \Phi^-(\O).\ee They satisfy the following commutation relation
\be
\left[
\hat{\Phi}^+(\Omega),
\hat{\Phi}^-(\Omega')
\right]
={8i  \over \sqrt{h}}\delta^3(\O-\O').
\ee We may then decompose these \ip\  operators as the sum of two terms
\be \label{dak} \hat \Phi^\pm(\O)=\hat \Phi^\pm_N(\O)+\hat \Phi^\pm_S(\O) \ee
where the first (second) acts only on $R^3_N$ ($R^3_S$). Defining the northern and southern Dirichlet and Neumann vacua by
\be \hat \Phi^\pm_N|0^\pm_N\>=0,~~~~\hat \Phi^\pm_S|0^\pm_S>=0, \ee
it follows from the  decomposition (\ref{dak}) that the global vacua have a simple product decomposition\footnote{Of course a general quantum state on $S^3$ is a sum of products of northern and southern states, and reduces to a southern density matrix, not a pure state,  after tracing over the northern Hilbert space. We shall see this explicitly below for the case of  the Euclidean vacuum.}
\be |0^\pm\>=|0^\pm_N\>|0^\pm_S\>.\ee
Excited southern states may then be built by acting on one of these southern vacua with
$\hat \Phi_S$. We wish to identify these states with those of the CFT$_3$ on $S^2$.

In the higher-spin dS/CFT correspondence there are actually two CFT$_3$'s living on \ip: the free $Sp(N)$ model, associated to
Neumann boundary conditions, and the critical $Sp(N)$ model, associated to Dirichlet boundary conditions. Since the field operators $\hat \Phi_S $ acting on  $|0_S^+\>$ ($|0_S^-\>$)  obeys, according to equation (\ref{bxz}),  Neumann (Dirichlet) boundary conditions near the southern hemisphere of \ip, it is natural to identify
\bea  |0_S^+\>&\sim& {\rm free}~~Sp(N)~~{\rm vacuum}\cr |0_S^-\>&\sim& {\rm critical}~~Sp(N)~~{\rm vacuum}.\eea

Next we want to consider excited states and their duals. To be specific we consider
the Neumann theory built on $|0_S^+\>$. Parallel formulae apply to the Dirichlet case.
Operator versions of the classical wavefunctions  $\Phi^-_\O(x)$ are constructed as
\be \hat\Phi^-_{\O_S} \equiv \left\langle \Phi^-_{\O_S}|\hat  \Phi\right\rangle_{R_S^3},\ee
where $\O_S$ is presumed to lie on the southern hemisphere.
We can make a quantum state
\be \label{dxl}|\O^-_S\>\equiv \hat\Phi^-_{\O_S} |0^+_S \rangle =\hat\Phi^-(\O_S)|0^+_S \rangle , \ee
where in the last line we used  (\ref{fc}).
By construction this will be a lowest weight state, and we therefore identify it as the bulk dual
to the CFT$_3$ state created by the primary operator $\co$ dual to the field $\Phi$.

This connection leads to an interesting nonperturbative dS exclusion principle \cite{mkss}. The operator $\co$ has a representation in the $Sp(N)$ theory as
%\be \co=\Omega_{AB}\chi^A\chi^B,~~~A,B=1,...N,\ee
\be \co=\Omega_{AB}\eta^A\eta^B,~~~A,B=1,...N,\ee
where $\eta^A$ are $N$ anticommuting real scalars and $\Omega_{AB}$ is the quadratic form on $Sp(N)$.  It follows that
\be \co^{{N\over 2} +1}=0.\ee
Bulk-boundary duality and the state-operator relation described above  then implies the nonperturbative relation
\be \left[ \hat \Phi^\pm(\Omega)\right]^{{N \over 2}+1}=0.\ee
Hence the quantum field operators  $\hat \Phi^\pm(\Omega)$ are $N \over 2$-adic. One is not allowed to put more than $N \over 2$ quanta in any given quasinormal mode. This is similar  to the stringy exclusion principle for AdS \cite{jmas} and may be related to the finiteness of dS entropy. Nonperturbative phenomena due to related finite $N$ effects in the $O(N)$ case have been discussed in \cite{Shenker:2011zf}. We hope to investigate further the consequences of this dS exclusion principle.

\subsection{Norm}
\label{normsubsection}
Having identified the bulk duals of the boundary CFT$_3$ states, we wish to describe the bulk dual of the CFT$_3$ norm.
The standard bulk  norm  is defined by $\Phi(x)=\Phi^\dagger(x)$. However  this norm is not unique. It has been argued for a variety of reasons beginning in \cite{Witten:2001kn} that it is appropriate to modify the norm in the context of dS -- see also \cite{Bousso:2001mw,Parikh:2004wh}. Here we have the additional problem that this standard norm is divergent for states of the form (\ref{dxl}).  We now construct the  modified norm for states on $R_3^S$ by demanding that it is equivalent to the CFT$_3$ norm. The construction here generalizes to dS$_4$ the one given in \cite{Bousso:2001mw} for dS$_3$.

The bulk action of dS Killing vectors $K_A^\mu\partial_\mu$ on a scalar field is generated by the integral over any global  $S^3$ slice
\be \hat {\cal L}_{A}= \int_{S^3}  d^3\Sigma^\mu T_{\mu \nu}K^{\nu}_A,\ee
where $T_{\mu\nu}$ is the bulk stress tensor constructed from the operator $\hat \Phi$. If we take $\hat \Phi^\dagger(x)=\hat \Phi(x)$, then $\hat {\cal L}_A=\hat {\cal L}_{A}^\dagger$ which is not what we want. The CFT$_3$ states are in representations of the $SO(3,2)$ conformal group. These arise from analytic continuation of the 10 $SO(4,1)$ conformal Killing vectors  on $S^3$ which
are the boundary restrictions of the bulk dS$_4$  Killing vectors $K^{\mu}_A \p_\mu$. Usually, the standard CFT$_3$ norm has a self-adjoint dilation operator ${\cal L}_0$ generating $-i \sin\psi\p_\psi$ as well as 3 self-adjoint $SO(3)$ rotation operators ${\cal J}_k$. The remaining 6 raising and lowering operators ${\cal L}_{\pm k}$ arising from the Killing vectors $i M_{\pm k}$ (described in the appendix) then obey ${\cal L}_{\pm k}^\dagger= {\cal L}_{\mp k}$  in the conventional CFT$_3$ norm.
%
% In the $Sp(N)$ models, however, the stress tensor on the Lorentzian cylinder is not hermitian with respect to the conventional norm. It is pseudo-hermitian,\footnote{
%We will define pseudo-hermitian and the operator $C$ in section \ref{pseudounitarity}.}  i.e. $T_{\mu\nu} =C T_{\mu\nu}^\dag C$ or $T_{\mu\nu}^{\dag_c} = T_{\mu\nu}$ if we defined $A^{\dag_c} \equiv C A^\dag C$ for any operator $A$. Thus, the appropriate adjoint action of $SO(3,2)$ that we want is ${\cal L}_0^{\dag_c}={\cal L}_0,~{\cal J}_k^{\dag_c}={\cal J}_k$ and ${\cal L}_{\pm k}^{\dag_c} = {\cal L}_{\mp k}$.

To obtain an adjoint with the desired properties, we define the modified adjoint
\be\label{ad} \hat  \Phi^{ \dagger}(x)= \calr \hat\Phi(x) \calr=\hat \Phi( R x) ,\ee where here and hereafter $\dagger$ denotes the bulk modified adjoint. The reflection operator $\calr$ is the discrete isometry of $S^3$ which reflects through the $S^2$ equator $R(\psi,\theta,\phi)=(\pi-\psi,\theta, \phi)$ along with complex conjugation. In particular, it maps the south pole to the north pole while keeping the equator invariant. This implies that ${\cal L}_0$ (generating $i L_0$) and ${\cal J}_k$ are self adjoint while
\be {\cal L}_{\pm k}^\dagger=-i\int_{S^3}  d\Sigma^\mu(x) T_{\mu \nu}(R x) M^{\nu}_{\pm k}(x)=-i\int_{S^3}  d\Sigma^\mu(x) T_{\mu \nu}(x)  M^{\nu}_{\pm k}(R x) = {\cal L}_{\mp k}.\ee
Hence we have constructed an adjoint admitting the desired $SO(3,2)$ action. We do not know whether or not it is unique.

The action of $\calr$ maps an operator defined on the southern hemisphere to one defined on the southern hemisphere of \ip\ according to
\be \label{dxw}\hat \Phi^{\pm\dagger}(\O)= \hat \Phi^\pm (\O_R),\ee
Hence the action of $\calr$ exchanges the northern and southern hemispheres, and maps a southern \ip\ state to a northern one.
Therefore it cannot on its own define an adjoint within the southern Hilbert space. For this we must combine (\ref{ad}) with a map from the north to the south. Such a map is provided by the Euclidean vacuum. The global Euclidean  bra state (constructed with the standard adjoint) can be decomposed in terms of a basis of northern and southern bra states
\be \<0_E|=\sum_{m,n}E_{mn}\< m_S|\<n_N|.\ee
We then define the modified  adjoint of an arbitrary  southern state $|\Psi_S\>$ by
\be|\Psi_S\>^\dagger \equiv \<0_E| \calr |\Psi_S\>.\ee
We will denote the corresponding inner product by an $S$ subscript
\be \<\Psi'_S|\Psi_S\>_S\equiv (|\Psi'_S\>^\dagger)|\Psi_S\>.\ee
For example choosing the basis so that
\be \calr |m_S\>=|m_N\> \ee
we have
\be \<m_S|n_S\>_S= E_{nm}
.\ee
In particular one finds
%\be |0^+_S\>^\dagger=\<0^-_S|e^{-\half \sum (-)^L a^+_{S,Lj}a^+_{S,Lj}},\ee
%so that with this norm
%\be ||0^+_S\>|^2= N_0^{-1}\<0_E|(|0^-_S\>\cp\ct |0^-_S\>)=1.\ee
\be \<0^+_S|0^+_S\>_S= \<0_E|(|0^+_S\>\calr |0^+_S\>)=1.\ee

Let us now compute the norm of the southern state $|\O^-_S\>$ in (\ref{dxl}). The action of $\calr$ gives a northern state which we will denote $|R \O^-_S\>$. The norm is
then
\be \<\O^-_S|\O^-_S\>_S
= \<0_E|(|\O^-_S\>|R\O^-_S\>)
= \<0_E|\hat \Phi^-(\O_S)\hat \Phi^-(R\O_S)|0^+\>. \ee
Using the relation
\be |0_E>=N_0
e^{-{1\over 16} \int d^3\O d^3\O'\hat \Phi^+(\O)\Delta_-(\O,\O')\hat \Phi^+(\O')}|0^-\rangle   \ee
we find
\be
\label{fidk} \<\O^-_S|\O^-_S\>_S=8\Delta_-(\O_S,R \O_S).
\ee
This is proportional to the $S^3$ two-point function of a dimension $h_-$ primary at the points $\O_S$ and $R \O_S$. The analogous computation in the
Dirichlet theory gives \be
\label{fik} \<\O^+_S|\O^+_S\>_S=-8\Delta_+(\O_S,R \O_S).
\ee
%with the relative minus sign traced to the one in (\ref{dxw}) for the adjoint of $\hat \Phi^\pm$.
\section{ Boundary  dual of the bulk Euclidean  vacuum}
In the preceding section we have argued that dS/CFT  maps CFT$_3$ states on an $S^2$ in \ip\ to bulk states on the southern slice ending on the $S^2$. A generic state in a global dS slice does not restrict to  a pure southern state. However we can always define a density matrix
by tracing over the northern Hilbert space. In particular, such a southern density matrix $\rho^E_S$ can be associated to the global Euclidean vacuum $|0_E\>$. The choice of an equatorial $S^2$ in \ip\ breaks the $SO(4,1)$ symmetry group down to $SO(3,1)$, which  also preserves the hyperbolic slices ending on the $S^2$.  $\rho_E^S$ must be invariant under this $SO(3,1)$. In fact $\rho_E^S$ follows from results in \cite{Sasaki:1994yt}. Writing the quadratic Casimir of $SO(3,1)$ as $C_2=-(1+p^2)$, it was shown, in a basis which diagonalizes $p$, that
\be \rho_E^S=N_1 e^{-2\pi p},\ee
where $N_1$ is determined by $\tr \rho_E^S=1$.  It would be interesting to investigate this further and compute the entropy $S=-\tr \rho_E^S\ln \rho_E^S$ in the $Sp(N)$ model.

\section{Pseudounitarity and the C-norm in the $Sp(N)$ CFT$_3$ }
\label{pseudounitarity}
In this section we consider the $Sp(N)$ model (where $N$ is even) and compare the norms to those computed above.  The action   is
\be
\label{dcf}
I_{Sp(N)}=\frac{1}{8\pi}
\int d^3 x
\left[
\delta^{ij}\delta_{ab}\partial_i \bar{\chi}^a \partial_j \chi^b
+m^2 \bar{\chi} \chi
+\lambda \left( \bar{\chi} \chi\right)^2
\right],
\ee where
$\chi^a (a=1,\ldots,{ N \over 2})$  is a complex anticommuting scalar and $\bar{\chi}\chi \equiv \delta_{ab} \bar{\chi}^a\chi^b$.
 This has a global  $Sp(N)$ symmetry and we restrict to $Sp(N)$ singlet operators.\footnote{The $U({N \over 2})$ theory has the same action but is restricted only to  $U({N \over 2})$ singlets.} For the free theory $m=\lambda=0$ while the critical theory is obtained by flowing to a nontrivial fixed point $\lambda_F$.
The $Sp(N)$ theory is not unitary in the sense that in the standard  norm following from (\ref{dcf}) one has that \cite{LeClair:2007iy}
\be H\neq H^\dagger \ee
 and $\langle \Psi'|\Psi \rangle$ is not preserved. Nevertheless, as detailed in \cite{LeClair:2007iy}, there exists an operator $C$ with the properties
 \be \label{cn} C^\dagger C=C^2=1,~~~C\chi^\dagger C=\overline{\chi}, ~~~CH^\dagger C=H,~~~C|0\>=|0\>.\ee  To write it in real fields, for e.g., in the case of $Sp(2)$, writing the real and imaginary part of $\chi$ as $\eta_1$ and $\eta_2$, the action of $C$ becomes $\eta_2 = C \eta_1^\dagger C$.
One may then define a ``pseudounitary" $C$-inner product \be  \< \Psi'|\Psi\>_C\equiv \<\Psi'|C|\Psi\>\ee
which is preserved under hamiltonian time evolution.  Such hamiltonians are pseudohermitian and are similar to those studied in \cite{Bender}. We note that the norm is not positive definite.

 Inserting an operator $O_i$ constructed from $\chi^a$ at the south pole gives a functional of the boundary conditions on the equatorial $S^2$ which we define as the state $|O_i \rangle$. This is the standard state-operator correspondence.
An inner product for  such states associated to $O_i$ and $O_j$ can be defined by the two point function with $O_i$ at one pole and $O_j$ at the other. It follows from (\ref{cn}) that this
is the $C$-inner product for the states $|O_i\rangle$ and $|O_j\rangle$:
\be
\<O_i|O_j\>_C= \<O_i|C|O_j\> = \<{O_i}^\dagger C {O_j}\>
=\<O_i O_j\>.
\ee In the last line, we used the fact that the (singlet) currents in the $Sp(N)$ models satisfy $
C {O_i}^\dagger C=O_i$ since $ C\left(\bar{\chi}\chi\right)^{\dag} C=\bar{\chi}\chi $.  %As discussed in subsection \ref{normsubsection}, we have ${\cal L}_{\pm k}^{\dag_c} = {\cal L}_{\mp k}$, from which it follows that
%\be
%\<{\cal L}_{-k}O_i|{\cal L}_{-k}O_j\>_C
%= \<O_i|C {\cal L}_{-k}^{\dag_c} {\cal L}_{-k} |O_j\> =
%2 h_i\<{O_i} {O_j}\>
%\ee which is just the usual Zamolodchikov norm for the first descendant.
% For a general descendant of $O_i$, which is of the form ${\cal L}_{-k_1}\ldots {\cal L}_{-k_n}  |O_i \rangle$, the above argument carries through in the same way. Thus, the C-norm for a primary state and its descendants equals to he Zamolodchikov norm.
%
For primary operators of weight $h_i$ we then have \cite{Anninos:2011ui}
\be \<O_i|O_i\>_C=-N\Delta_{h_i}(\O_{NP},\O_{SP}). \ee Hence it is the C-norm which maps under the state-operator correspondence to the Zamolodchikov norm defined as the Euclidean two point function on $S^3$. As seen in \cite{Anninos:2011ui}
this C-norm then agrees  with the bulk inner product (\ref{fidk})-(\ref{fik}) of the dual state for the scalar case.\footnote {A factor of $i^h$ explained in \cite{Anninos:2011ui} relating operator insertions to bulk fields makes the two-point function (\ref{fidk}) negative.  } Moreover, as the bulk  and CFT$_3$ norms assign the same hermiticity properties to the $SO(4,1)$ generators, this result will carry over to descendants of the primaries. A generalization of this construction to all spins seems possible.

One of the puzzling features of dS/CFT is that the dual CFT  cannot be unitary in the ordinary sense. This is not a  contradiction of any kind because unitarity of the Euclidean CFT is not directly connected to any spacetime conservation law. At the same time quantum gravity in dS -- and its holographic dual -- should have some good property replacing unitarity in the AdS case. It is not clear what that good property is.  The appearance of a pseudounitary structure in the case of dS/CFT analyzed here is perhaps relevant in this regard.

\section*{Acknowledgements}
It has been a great pleasure discussing this work with Dionysios Anninos, Daniel Harlow, Tom Hartman, Daniel Jafferis,  Matt Kleban and Steve Shenker. This work was supported in part by DOE grant DE-FG02-91ER40654 and the Fundamental Laws Initiative at Harvard.

\appendix

\section{Appendix: dS$_4$ Killing vectors}
\label{appendix1}
The 10 Killing vectors of dS$_4$ are given by:
\bea
L_0&=&\cos \psi \p_t-\tanh t \sin \psi \p_\psi\nonumber\\
M_{\mp 1}&=&
\pm
\sin{\psi}\sin{\theta}\sin{\phi} \partial_t
 + \left( 1 \pm \tanh{t} \cos{\psi}\right)\sin{\theta}\sin{\phi}\partial_\psi \nonumber\\
& &
+\left( \cot{\psi} \pm \tanh{t} \csc{\psi} \right)\left(
\cos{\theta}\sin{\phi}\partial_\theta
+\csc{\theta}\cos{\phi}  \partial_\phi\right)
 \nonumber\\
 M_{\mp2} &=&
 \pm \sin{\psi} \sin{\theta} \cos{\phi}\partial_t
 + \left( 1 \pm \tanh{t} \cos{\psi}\right)\sin{\theta}\cos{\phi}\partial_\psi
 \nonumber\\
 & & +\left( \cot{\psi} \pm \tanh{t} \csc{\psi}\right) \left(
 \cos{\theta}\cos{\phi}\partial_\theta
-\csc{\theta}\sin{\phi} \partial_\phi \right)\nonumber\\
M_{\mp3}
&=& \pm \sin{\psi} \cos{\theta}  \partial_t
+ \left(1\pm \tanh{t}\cos{\psi} \right) \cos{\theta}\partial_\psi
-\left(\cot{\psi} \pm \tanh{t} \csc{\psi}\right)\sin{\theta} \partial_\theta\nonumber\\
J_1 &=& \cos{\phi}\partial_\theta -\sin{\phi}\cot{\theta}\partial_\phi\nonumber\\
J_2 &=& -\sin{\phi}\partial_\theta -\cos{\phi}\cot{\theta}\partial_\phi\nonumber\\
J_3 &=& \partial_\phi.
\eea

For each $k$, the $M_{\pm k}$ and $L_0$ form a $SO(2,1)$ subalgebra satisfying
\be
[M_{+k},M_{-k}]=2L_0,~~
[L_0,M_{+k}]=-M_{+k},~~
[L_0,M_{-k}]=M_{-k}.~~
\ee
As mentioned in the text, acting on $SO(3)$-invariant functions, we have
\be \ell^2\nabla^2=-L_0(L_0-3) +M_{-k}M_{+k},\ee where $k$ is summed over $k=1,2,3$.

The conformal Killing vectors of the $S^3$ are given by the restriction of
dS$_4$ Killing vectors on \ip:
\bea
L_0&=& -\sin{\psi} \partial_\psi \nonumber\\
M_{\mp 1}&=&
\left( 1 \pm  \cos{\psi}\right)\sin{\theta}\sin{\phi}\partial_\psi
+\left( \cot{\psi} \pm  \csc{\psi} \right)\left(
\cos{\theta}\sin{\phi}\partial_\theta
+\csc{\theta}\cos{\phi}  \partial_\phi\right)
 \nonumber\\
 M_{\mp2} &=&  \left( 1 \pm  \cos{\psi}\right)\sin{\theta}\cos{\phi}\partial_\psi
+ \left( \cot{\psi} \pm  \csc{\psi}\right) \left(
 \cos{\theta}\cos{\phi}\partial_\theta
-\csc{\theta}\sin{\phi} \partial_\phi \right)\nonumber\\
M_{\mp3}
&=&  \left(1\pm \cos{\psi} \right) \cos{\theta}\partial_\psi
-\left(\cot{\psi} \pm \csc{\psi}\right)\sin{\theta} \partial_\theta.\nonumber\\
J_1 &=& \cos{\phi}\partial_\theta -\sin{\phi}\cot{\theta}\partial_\phi\nonumber\\
J_2 &=& -\sin{\phi}\partial_\theta -\cos{\phi}\cot{\theta}\partial_\phi\nonumber\\
J_3 &=& \partial_\phi.
\eea

\end{document}